# Theory on the mechanism of rapid binding of transcription factor proteins at specific-sites on DNA

*Rajamanickam Murugan*[*]
*Department of Biotechnology, Indian Institute of Technology Madras*
*Chennai 600036, India*

[*] rmurugan@gmail.com

ABSTRACT
We develop revised theoretical ideas on the mechanism by which the transcription factor proteins locate their specific binding sites on DNA faster than the three-dimensional (3D) diffusion controlled rate limit. We demonstrate that the 3D-diffusion controlled rate limit can be enhanced when the protein molecule reads several possible binding stretches of the template DNA via one-dimensional (1D) diffusion upon each 3D-diffusion mediated collision or nonspecific binding event. The overall enhancement of site-specific association rate is directly proportional to the maximum possible sliding length ($L_A$, square root of $(6D_o/k_r)$ where $D_o$ is the 1D-diffusion coefficient and $k_r$ is the dissociation rate constant associated with the nonspecific DNA-protein complex) associated with the 1D-diffusion of protein molecule along DNA. Upon considering several possible mechanisms we find that the DNA binding proteins can efficiently locate their cognate sites on DNA by switching across fast-moving, slow-moving and reading states of their DNA binding domains in a cyclic manner. Irrespective of the type of mechanism the overall rate enhancement factor asymptotically approaches a limiting value which is directly proportional to $L_A$ as the total length of DNA that contains the cognate site increases. These results are consistent with the *in vitro* experimental observations.

## INTRODUCTION

Binding of transcription factor (TF) proteins at *cis*-regulatory specific-sites located on the genomic DNA is essential to activate as well as regulate the transcription of various genes across prokaryotes to eukaryotes (**1-2**). The site-specific interactions of protein molecules with genomic DNA were considered earlier as one-step three-dimensional (3D-only model) diffusion controlled collision processes. Later *in vitro* experimental studies (**3-4**) on site-specific binding of *lac* repressor protein with its Operator sequence located on a template DNA showed a bimolecular site-specific collision rate in the order of $\sim 10^{10}$ $mol^{-1}$ $s^{-1}$ which is $\sim 10^2$ times faster than that of the 3D-only diffusion controlled collision rate limit in aqueous medium. This result and subsequent studies revealed that the underlying dynamics might be oversimplified by the 3D-only diffusion models and suggested that such higher bimolecular collision rates could be achieved when the mechanism of searching of protein molecules for their specific binding sites on DNA is through a combination of 3D excursions and one-dimensional (1D) diffusion along the genomic DNA (3D1D model) (**3-6**).

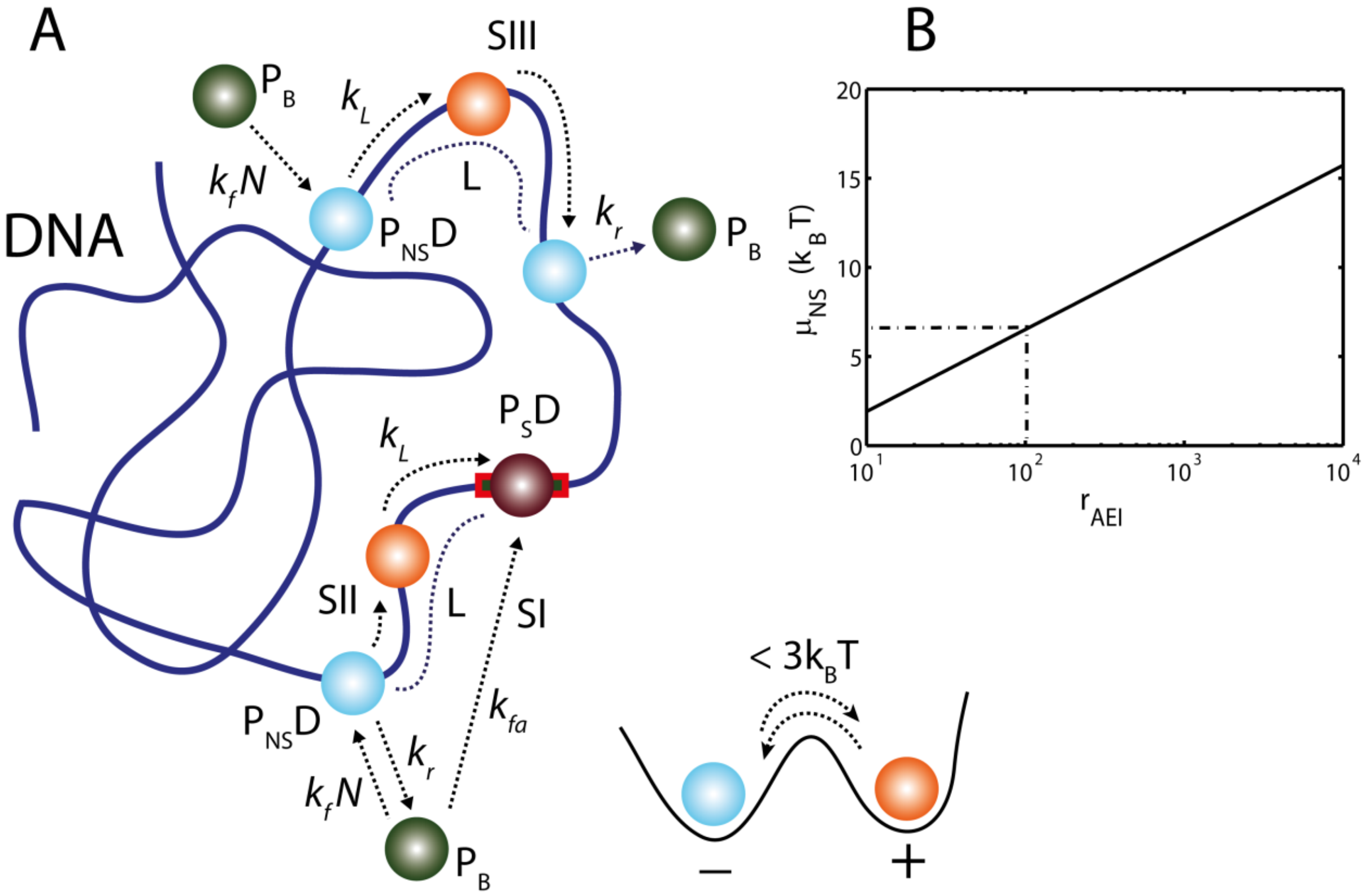

**FIGURE 1**. **A**. Various models on site-specific DNA protein interactions are shown. In the scheme **SI**, the protein molecule from bulk ($P_B$) directly binds with its specific site to form specific complex ($P_SD$) via 3D diffusion mediated collision where the maximum achievable rate limit will be $k_{fa} \sim 10^8$ $M^{-1}s^{-1}$ under *in vitro* condition which is the same as 3D-diffusion limited rate limit. In scheme **SII**, the protein molecule first nonspecifically binds with DNA to form a nonspecific complex ($P_{NS}D$) with an on-rate limit of $\sim k_fN$ and an off-rate of $k_r$ ($s^{-1}$) and then reaches the specific site without dissociation via 1D diffusion along the DNA chain with a rate $k_L$ that depends on the distance ($L$, the contour length of DNA segment that intervenes the initial landing position of protein and its specific binding site) of the specific site from the initial landing position. Here we have $k_f \sim 10^8$ $M^{-1}s^{-1}bps^{-1}$ under *in vitro* conditions where $N$ ($\sim 4.6 \times 10^6$ bps for *E. coli* genome) is the size of the template DNA. In scheme **SIII**, contrasting from **SII** the protein molecule undergoes several dissociations after reading $L$ number of possible binding stretches of DNA before reaching the specific binding site. In the process of searching, the DNA binding domains of DNA binding proteins undergo conformational fluctuations between two different states namely fast-moving (+ve) and slow-moving (-ve). Fast-

moving state is lesser sensitive to the sequence information than the slow-moving states. The free energy barrier that separates these two states seems to be in the range of ~ (1-3) $k_BT$ similar to that of the folding-unfolding dynamics of downhill folding proteins at their mid-point denaturant temperatures. **B**. Dependency of the free energy barrier associated with the dissociation of $P_{NS}D$ on the rate enhancement factor over 3D-diffusion controlled rate limit $r_{AEI}$ (measured in terms of bps) as given by **Eq. 8**. Here $D_o$ ~ $10^6$ $bps^2s^{-1}$ and $k_r^0$ ~$10^6s^{-1}$ and for the experimentally observed $r_{AEI}$ ~ 100 bps, we find $\mu_{NS}$ ~ $6k_BT$.

According to the current literature (**3-20**), the 3D diffusion mediated site-specific interactions of proteins with DNA seem to be facilitated by several other concurrent dynamical processes such as 1D sliding, hopping and inter-segmental transfers. In sliding mode, the non-specifically bound protein molecule diffuses along the DNA chain with unit base-pair (bps) step size whereas it can leap over few bps at a time in the hopping-mode. Sliding and hopping modes of 1D dynamics dominate on linear and loosely packaged segments of genomic DNA. When the DNA polymer is condensed and super-coiled, then the diffusing protein molecule can undergo inter-segmental transfers via ring closure events which can occur whenever two distal segments of the same DNA chain come into contact in 3D space upon condensation (**Fig. 1A**). Here one should note that these facilitating processes ultimately reduce the overall search-time required by the protein molecule to locate its specific site on DNA mainly by fine-tuning the ratio of search-times spent on 1D and 3D diffusion routes. It seems that the overall search-time can be minimized when the protein molecule spends equal amount of times both in 1D and 3D routes (**7-8**). This means that there exists an optimum sliding-length ($L_C$ measured in base-pairs) at which the overall search time is a minimum ($\tau_S$). Detailed theoretical studies as well as the single molecule experiments (7-**15**) substantiated these ideas and further suggested that the spatial organization and packaging (**10-11**, **19**) of the DNA molecule can significantly enhance the rate of site-specific DNA-protein interactions. Apart from these the DNA binding domains (DBDs) of the non-specifically bound DNA binding proteins (DBP) seems to thermally fluctuate between at least two different conformations namely the ‘fast-moving' state and 'slow-moving' states. Upon finding the specific-sites, these thermally driven conformational fluctuations in the DBDs of DBPs are damped out that result in the formation of a tight site-specific DNA-protein complex (**20-21**). Recent theoretical investigations (**21**) suggested that the conformational flipping of DBDs of DBPs must resemble that of the dynamics of downhill folding proteins at their mid-point ($T_m$) denaturation temperatures (**21**). It seems that an efficient thermodynamic coupling between the conformational flipping with the 1D search dynamics of DBPs can be achieved (**21**) when the free energy barrier that separates the fast- and slow-moving states of DBDs is ~$k_BT\ln 2$. It seems that at this optimum barrier height both the speed of searching and accuracy in the detection of specific-sites by the DBDs of DBPs will be a maximum.

Mechanism of site-specific DNA-protein interactions have been revisited by several groups recently (**16-18**). Most of the studies assessed the 3D1D-model on the basis that the 1D diffusion is always slower than the 3D diffusion and therefore it cannot enhance the overall bimolecular site specific association rate. Other studies suggested (**22**) that the presence of electrostatic attractive potential between DBDs of DBPs and DNA backbone may be enough to enhance the 3D diffusion controlled collision rate which means that the 3D-only model is enough to describe the site specific DNA-protein interactions. More recently Zhou (**23**) has questioned the validity of the existence of optimum search time and optimum sliding length since the presence of optimum sliding length further predicts the existence of an optimum length of template DNA at which the overall bimolecular association rate will be a maximum. Several inconsistences of such 1D random walk models in explaining the experimentally observed (**23**) monotonic variation of the bimolecular association rate with the length of DNA were pointed out and consecutively a rigorous approach was proposed (**23**) in which the template DNA polymer was modelled as a cylinder immersed in aqueous medium and the problem of site-specific DNA-protein interactions was described as a two-dimensional (2D) diffusion of protein molecules on the cylindrical surface of DNA. Eventually in these models the protein searches for its specific-binding patch on the surface of the template DNA-cylinder through a combination of 2D and 3D diffusion and the overall problem was formulated in cylindrical coordinate system. One should note that such approaches completely ignore the three dimensional coiled conformation of DNA and as a consequence they will underestimate the overall site-specific association rate. In this paper, we show that the 3D1D models are indeed more accurate in predicting the overall bimolecular site-specific association rate and its monotonic dependency on the length of template DNA and various other experimental observations.

## THEORY

Let us consider a template DNA molecule with size of *N* base-pairs (bps) that contains a specific binding site for an arbitrary transcription factor protein (TF). Under aqueous conditions, the dissolved DNA polymer takes a random coil conformation and we denote the molar concentration of entire DNA as [D] (measured in mol/lit, M). This means that the concentration of specific binding site will be the same as [D]. Noting the fact that a frame shift of single base-pair can result in a nonspecific binding site for the corresponding TF protein, one can conclude that the total concentration of non-specific binding sites will be on the order of ($N$-$m$) [D] ~ $N$ [D] under homogenous conditions where $m$ is the number of base-pairs of DNA covered upon binding of the TF protein. Let us denote the concentration of TF protein as [P] which has three possible components under equilibrium conditions namely protein in bulk [$P_B$], protein molecules which are non-specifically bound with DNA [$P_{NS}$] and protein molecules which are site specifically bound with DNA [$P_S$]. The total concentration of TF protein will be [$P_T$] = [$P_B$] + [$P_S$] + [$P_{NS}$]. We will denote the nonspecific protein-DNA complex as $P_{NS}D$ and specific DNA-protein complex as $P_SD$. With these settings, the TF protein molecules can bind with their specific sites on DNA via two possible reaction mechanisms (**23**) as depicted by the following schemes SI and SII.

$$P_B + D \xrightarrow{k_{fa}} P_S D \text{ (SI)}; \; P_B + D \underset{k_r}{\overset{k_f}{\rightleftarrows}} P_{NS}D \xrightarrow{k_L} P_S D \text{ (SII)}$$

Scheme SI describes the one step three dimensional (3D) diffusion mediated mechanism with a bimolecular collision rate limit of $k_{fa} \sim 10^8 M^{-1}s^{-1}$ under *in vitro* conditions. However several *in vitro* experiments revealed a bimolecular association rate constant of the order ~$10^9$ to $10^{10}$ $M^{-1}s^{-1}$ that is 10-100 times faster than the 3D diffusion controlled rate limit henceforth ruled out the possibility of the mechanism given in **SI**. There are several counterarguments to this deduction. One of them is that the electrostatic interactions at the DNA-protein interface could contribute to the rate enhancements over the diffusion controlled collision rate limit (**22**). However one should note that such electrostatic forces can act only on short distances compared to the 3D diffusion length of the protein molecule under aqueous conditions that in turn weakens such arguments as shown by Kolomeisky in reference (**16**). In the second scheme SII, the TF protein first nonspecifically binds with DNA with a bimolecular rate limit of $k_f \sim N10^8$ $M^{-1}s^{-1}bps^{-1}$ and then it searches for the specific site through a rate limiting one-dimensional (1D) diffusion step. The differential rate equations corresponding to **SII** can be written as follows.

$$d[P_{NS}D]/dt = k_f N[P_B][D] - (k_r + k_L)[P_{NS}D];\ d[P_S D]/dt = k_L[P_{NS}D] \quad (1)$$

Here $k_r$ ($s^{-1}$) is the dissociation rate constant connected with the nonspecifically bound protein molecules on DNA ($P_{NS}D$) and $k_L$ ($s^{-1}$) is the first order rate constant associated with the site-specific binding of $P_{NS}D$ via 1D diffusion where $L$ is the minimum 1D diffusion length that is required by the $P_{NS}D$ to locate the specific site on DNA to form a tight $P_SD$ complex. One should note that $L$ is directly proportional to the distance between the initial position of $P_{NS}D$ and the position of its specific binding site on DNA and it can lie anywhere in the interval (0, $N$) with equal probabilities (=1/$N$). In **Eq. 1**, the time scale associated with the first reaction (formation of $P_{NS}D$) is much faster than the second rate limiting step that results in the following steady state approximation.

$$d[P_{NS}D]/dt \approx 0;\ d[P_S D]/dt \approx k_A(L)[P_B][D];\ k_A(L) = k_L k_f N/(k_r + k_L) \quad (2)$$

Here $k_A(L)$ is the apparent association rate (measured in terms of $M^{-1}s^{-1}$) that is observed in the *in vitro* experiments which in turn depends on the distance between the position of initial non-specific contact and the position of specific binding site (**Fig. 1A**) which are all located on the same stretch of DNA. One can rewrite the definition of association rate as follows.

$$k_A(L) = k_f N\phi(L);\ \phi(L) = 1/(1 + k_r\tau_L);\ \tau_L = L^2/6D_o \quad (3)$$

In this equation $\tau_L$ is the average 1D diffusion time (**5,6,7**, **24-26**) required by the non-specifically bound protein molecule $P_{NS}$ to read a stretch of $L$ bps of template DNA that contains the specific binding site to form site specific $P_SD$ complex. Here $D_o$ ($bps^2s^{-1}$) is the 1D diffusion coefficient associated with the dynamics of $P_{NS}$ on the template DNA chain. Since we have $L \in (0, N)$ with equal probabilities (=1/$N$), the overall average value of the apparent site-specific association rate can be computed as follows.

$$\bar{k}_{AEI} = \int_0^N k_A(L)\,dL\Big/N = k_f L_A \arctan(N/L_A);\ L_A = \sqrt{6D_o/k_r} \quad (4)$$

In this equation $\partial_N \bar{k}_{AEI}$ will asymptotically approach zero whenever $N > L_A/\sqrt{3}$ since the third derivative $\partial_N^3 \bar{k}_{AEI}$ vanishes at $N_C = L_A/\sqrt{3}$. This means that the apparent association rate constant will be asymptotically insensitive to the increasing size of DNA and approaches a finite limit as $\lim_{N\to\infty} \bar{k}_{AEI} = (\pi/2) k_f L_A$ particularly whenever $N \gg N_C$ (as shown in **Fig. 2A** and **2B**). In the subscript 'AEI' in the definition of the overall rate constant $\bar{k}_{AEI}$, 'A' denotes the averaging procedure, 'E' denotes equal weights that are used for averaging and 'I' denotes the integration method. Using the expression for $\bar{k}_{AEI}$ ($M^{-1}s^{-1}$) one can define the number of times ($r_{AEI}$) the observed average apparent association rate is higher than that of the 3D diffusion controlled collision rate limit (we can also denote it as the rate enhancement factor) as follows.

$$r_{AEI} = \bar{k}_{AEI}/k_f;\ 0 \le r_{AEI} \le N;\ \lim_{N\to\infty} r_{AEI} = (\pi/2)L_A;\ \lim_{k_r\to\infty} r_{AEI} = 0;\ \lim_{k_r\to 0} r_{AEI} = N \quad (5)$$

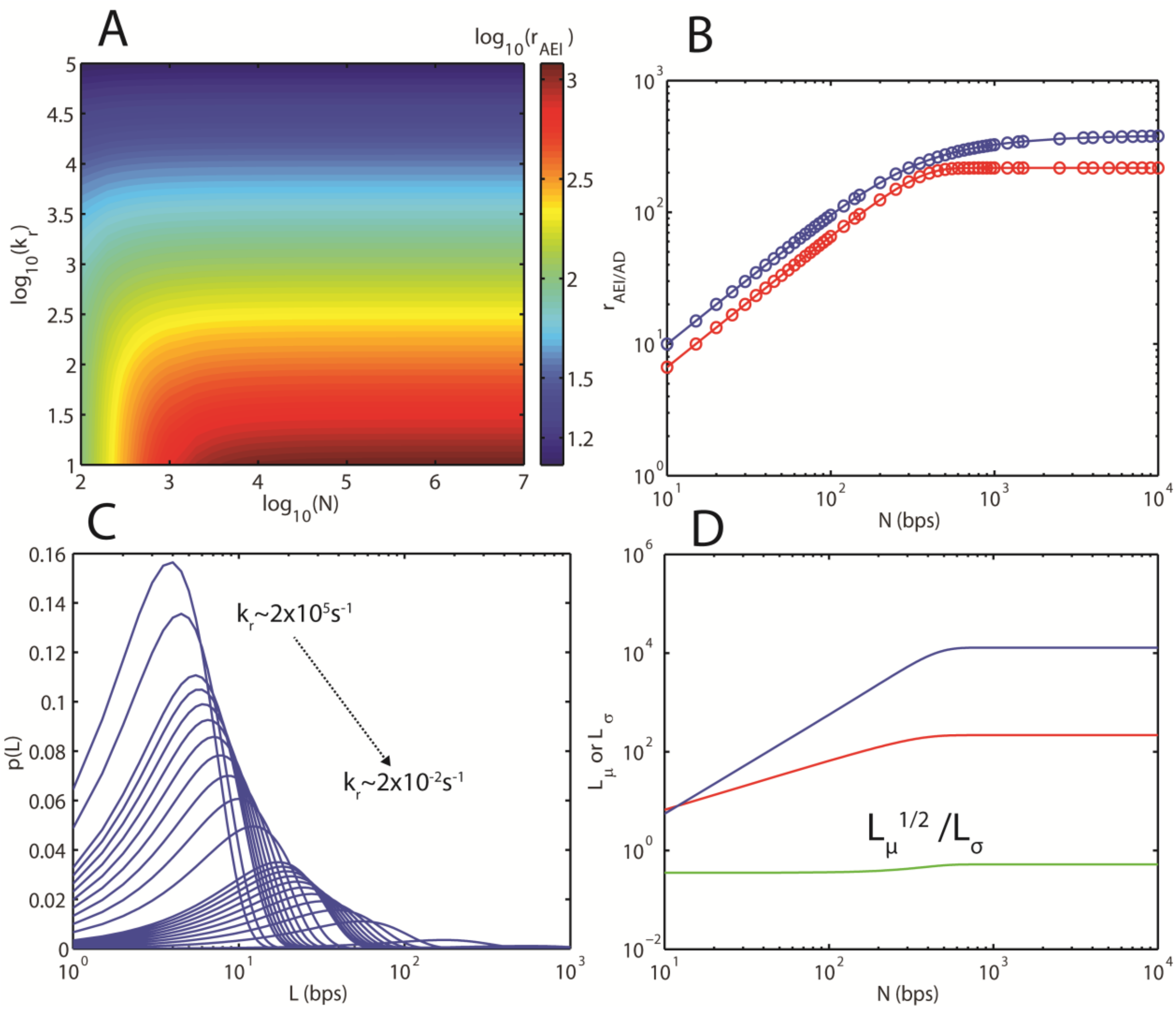

**FIGURE 2**. **A**. Variation of the overall rate enhancement factor that can be achieved in SII ($r_{AEI}$ bps) with respect to changes in the dissociation rate $k_r$ constant and size of the template DNA $N$. Here we have set the 1D diffusion-coefficient as $D_o$~$10^6$ bps$^2$s$^{-1}$ in **Eq. 5**. **B**. Asymptotic variation of rate overall rate enhancement factor in SII ($r_{AEI}$, blue solid line with circles) and SIII ($r_{AD}$, red solid line with circles) with respect to changes in the size of the template DNA $N$. Here we have set the 1D diffusion-coefficient as $D_o$~$10^6$ bps$^2$s$^{-1}$ and $k_r$~$10^2$s$^{-1}$ ( $L_A = \sqrt{6D_o/k_r}$ ) in **Eq. 5** and with these values to calculate $r_{AD}$ we have set $n$ = 25 and $\varepsilon$ = $10^{-5}$ in **Eq. 18**. At sufficiently large values of $N$, we find the asymptotic limiting values that $r_{AEI} = (\pi/2)L_A$~385 bps, $r_{AD} = \sqrt{\pi}\,L_A/2$~217 bps and clearly $r_{AEI}$ is an overestimate owing to the assumption that the nonspecific complex does not dissociates until it reaches its specific binding site. **C**. Variation of the probability density function $p(L)$ associated with the 1D reading lengths ($L$) in SIII with respect to changes in the dissociation rate constant $k_r$. Here $k_r$ is iterated from 2x$10^5$ to $10^{-2}$s$^{-1}$. **D**. Variation of mean (red solid line), variance (blue solid line) and coefficient of variation (green solid line) of reading lengths corresponding to $p(L)$ in **SIII** with respect to changes $N$. At higher values of $N$ we can obtain the asymptotic value of mean as $L_\mu = \sqrt{\pi}L_A/2$, variance as $L_\sigma = L_\mu^2(4-\pi)/4$ and coefficient of variation (CV) becomes as $\sqrt{L_\sigma}/L_\mu = \sqrt{(4-\pi)/\pi}$ . Upon comparison with **B** we find that the maximum achievable overall rate enhancement factor $\lim_{\varepsilon\to 0} r_{AD} = L_\mu$.

Since the bimolecular nonspecific collision rate $k_f$ is measured in terms of M$^{-1}$s$^{-1}$bps$^{-1}$ in the current context, the rate enhancement factor $r_{AEI}$ will be measured in terms of bps$^1$. In case of **Eqs. 4-5** we have assumed $L$ as a continuous type variable. When the distance $L$ is considered as a discrete type variable then we need to replace the integral by the corresponding discrete sum in these equations as follows.

$$\bar{k}_{AES} = \sum_{L=0}^{N} k_A(L)/N = k_f L_A \left(\Psi(N+1-\mathrm{i}L_A) - \Psi(-\mathrm{i}L_A) - \Psi(N+1+\mathrm{i}L_A) + \Psi(\mathrm{i}L_A)\right)/2\mathrm{i} \quad (6)$$

Here $\Psi(X) = d\ln\Gamma(X)/dX$ is the digamma function where $\Gamma(X)$ is the regular gamma function and $\mathrm{i} = \sqrt{-1}$. From this summation formula one can find that for large values of $N$ as well as dissociation constant $k_r$, the rate enhancement factor asymptotically approaches the following limits.

$$r_{AES} = \bar{k}_{AES}/k_f \; ; \; \lim_{N\to\infty} r_{AES} = \left(1+\pi L_A \coth(\pi L_A)\right)/2 \; ; \; \lim_{k_r\to\infty} r_{AES} = 1; \; \lim_{k_r\to 0} r_{AES} = N \quad (7)$$

From this equation one finds that the range of enhancement factor as $r_{AES} \in (1, N)$ (bps). Averaging over $L$ with equal weights (=1/$N$) as in **Eqs. 4** and **6** will be a valid procedure only when the concentration of $P_B$ is much lesser than the concentration of total nonspecific binding sites ~$N$ [D] so that there are limited number of protein molecules in the system which can land at all possible positions of the DNA polymer with equal probabilities which is true in most of the physiological or *in vitro* experimental conditions (**27-28**). Here one should note that **Eq. 6** predicts the possible range of enhancement factor reasonably better than **Eq. 4** particularly in the limit as $k_r \to \infty$. Most of the *in vitro* experiments on site specific DNA-protein interactions showed a rate enhancement of $r_{AEI/S} \sim 100$ bps irrespective of the size of the template DNA used in the experiment (**3-5**). When we define the rate constant associated with the dissociation of $P_{NS}D$ from the template DNA chain as $k_r = k_r^0 e^{-\mu_{NS}}$ where $\mu_{NS}$ is the free-energy barrier (measured in terms of $k_BT$ where $k_B$ is the Boltzmann constant and $T$ is the absolute temperature in $K$) associated with such dissociation phenomenon and $k_r^0$ is the maximum possible dissociation rate at zero free-energy barrier then from **Eqs. 5-6** in the limit as $N \to \infty$ we find the followings (**Fig. 1B**).

$$k_r = k_r^0 e^{-\mu_{NS}} = \left(3D_o\pi^2/2r_{AEI}^2\right); \;\; \therefore \mu_{NS} = -\ln\left(3D_o\pi^2/2k_r^0 r_{AEI}^2\right) \quad (8)$$

Noting the fact that $D_o \sim 10^6$ bps$^2$s$^{-1}$and $k_r^0 \sim 10^6$ s$^{-1}$ is the folding rate limit (**29-30**) we can find the probable value of the free-energy barrier connected with the dissociation of $P_{NS}D$ as $\mu_{NS}$ ~ 6 $k_BT$ corresponding to the experimentally observed rate enhancement factor of $r_{AEI}$ = 100 bps. In these calculations we have used the folding rate limit for $k_r^0$ due to the fact that the dissociation of $P_{NS}D$ essentially involves the segmental motion of DNA binding domains of TF proteins at their DNA-protein interface. These results suggest that the average residence time of $P_{NS}$ on DNA is (1/$k_r$) ~ 4ms and the required average 1D reading length of $P_{NS}D$ before dissociation from DNA or finding the specific site is given by $L_A$. In other words $L_A$ is the maximum number of possible binding stretches which can be read by the nonspecifically bound $P_{NS}$ before it dissociates from DNA and as a result we obtain $r_{AEI} \propto L_A$. Particularly for $r_{AEI} = 100$ we find the required value of this reading length as $L_A \approx 60$ bps. Undeniably this is how the rate of site specific interactions of TF proteins with DNA is enhanced over the 3D diffusion controlled rate limit. In case of pure 3D-diffusion route shown in scheme **SI**, the protein molecule reads only one binding stretch upon each 3D-diffusion mediated collision with the DNA chain whereas in case of two-step mechanism given in **SII** the protein molecule can read at most $L_A$ numbers of possible binding stretches upon each 3D collision with DNA with a rate of $k_f N$ that in turn leads to the proportionality relationship $r_{AEI} \propto L_A$.

So far we have assumed constant values for the 1D diffusion coefficient $D_o$, dissociation rate constant $k_r$ and nonspecific association constant $k_f$ throughout the DNA template. Though $k_f$ is almost independent of the sequence of DNA, $k_r$ is dependent on the DNA sequence since the free-energy barrier associated with the dissociation of $P_{NS}D$ ($\mu_{NS}$) is a sequence dependent quantity and the dissociation takes place against the non-specific electrostatic potential. One should note that $D_o$ will be independent of the DNA sequence since the 1D diffusion takes place across the electrostatic potential. Here one can consider the first order dissociation of $P_{NS}D$ as a rate process with a static disorder in the free-energy barrier (**31**). Under such conditions, $k_r$ will be a function of position ($s \in (0, N)$) of $P_{NS}D$ on the DNA sequence via the corresponding position dependent barrier height $\mu_{NS}(s)$ and as a consequence the apparent association rate constant will be a function of both $L$ and $s$ as follows.

$$k_{AS}(L,s) = k_f N \lambda(L,s);\ \lambda(L,s) = 1\big/\left(1 + k_r(s)\tau_L\right);\ \bar{k}_{AEIS} = \left(k_f/N\right)\int_0^N \int_0^N \lambda(L,s)\,dL\,ds \qquad (9)$$

In this equation the position dependent dissociation rate constant $k_r(s)$ takes random values depending on the guanine-cytosine (GC) content at a given binding position of $P_{NS}$ ($s$) on the DNA sequence. Here one should note that the double integral in **Eq. 9** cannot be calculated without sequence input. However to simplify the calculations we can consider the following limiting values of the apparent association rate constant.

$$\lim_{N\to\infty} \bar{k}_{AEIS} = (\pi/2) k_f L_A^0 \psi;\ \psi = \lim_{N\to\infty}\left\{\int_0^N e^{\mu_{NS}(s)} ds\Big/N\right\};\ L_A^0 = 6D_0\big/k_r^0 \qquad (10)$$

In this equation $\psi$ is the overall rate enhancing component since $L_A^0 \sim 1$. To evaluate the integral term in **Eq. 10** one can consider the position dependent free-energy barrier as a sum of a constant mean-value term and position dependent random term with the following average properties.

$$\mu_{NS}(s) = \bar{\mu}_{NS} + \xi_s;\ \langle \xi_s \rangle = 0;\ \langle \xi_s \xi_{s'} \rangle = \sigma_\mu^2 \delta(s - s');\ \xi_s \in N\left(0, \sigma_\mu^2\right) \qquad (11)$$

When the position dependent random free-energy $\xi_s$ is approximately distributed as Gaussian with zero mean, finite variance $\sigma_\mu^2$ and other zero higher order odd moments then we can find the expression for the rate enhancing component $\psi$ in **Eq. 10** as follows.

$$\psi = e^{\bar{\mu}_{NS}} \lim_{N\to\infty}\left\{\int_0^N e^{\xi_s} ds\Big/N\right\} \approx e^{\bar{\mu}_{NS}}\left(1 + \sigma_\mu^2\right);\ \because e^{\xi_s} = 1 + \xi_s + \xi_s\xi_s/2 + \ldots \qquad (12)$$

The approximation given in this equation will be valid only for small values of $\sigma_\mu^2$ and for higher values of the variance one needs to include higher order (even) moments in the calculation of $\psi$ . Using the expression of $\psi$ one can write the expression for limiting value of the overall apparent association rate constant as follows.

$$\lim_{N\to\infty} \bar{k}_{AEIS} \approx (\pi/2) k_f \bar{L}_A \left(1+\sigma_\mu^2\right);\ \bar{L}_A = 6D_0/\bar{k}_r;\ \bar{k}_r = k_r^0 e^{-\bar{\mu}_{NS}} \tag{13}$$

Here $\bar{L}_A$ is the overall average number of binding stretches read by the nonspecifically bound TF protein $P_{NS}$ before its dissociation. **Eq. 13** suggests that the extent of static disorder within the profile of free-energy barrier associated with the dissociation of $P_{NS}D$ also plays critical roles in enhancing the overall site-specific association rate. In deriving **Eqs. 1-13** we have assumed that (a) the nonspecific protein-DNA complex $P_{NS}D$ does not dissociate from DNA until it reaches the specific site via 1D diffusion from its initial landing position to form $P_SD$ and (b) the DNA binding domain (DBD) of the transcription factor protein is intact and does not fluctuate across various possible conformational states. Under *in vivo* conditions it seems that most of the DNA binding proteins (DBPs) always found to be nonspecifically bound with the genomic DNA as revealed by several experimental studies (**3-4**, **13-15**) which henceforward validate our first assumption (a). However in all these calculations we have assumed that the 1D reading lengths of TF proteins are lesser than the mean free path length. Mean free path length ($L_M$) is the average distance for which a given TF protein molecule can actually perform 1D-diffusion along DNA without collisions with other roadblock protein molecules ($P_R$) those are present on the same DNA chain. This means that the presence of roadblock protein molecules will alter the distribution profiles of $k_L$ and $k_r$ and **Eqs. 1-13** will be valid only when the total protein concentration ($[P_B]+[P_{NS}]+[P_R]$) is much lesser than the total concentration of the nonspecific binding sites $N$ [D]. Under crowded environments the most probable value of 1D reading length $L_M \sim N\ [D]/([P_B]+[P_{NS}]+[P_R])$ will be lesser than $L_A$ and our assumption in two-step reaction scheme **SII** that the nonspecifically bound TF protein molecule reaches the specific site via pure 1D diffusion without dissociation will not be true and as a consequence the process of averaging with equal weights in **Eq. 4** is not valid. On the other hand, under such conditions one can compute the average time required by the TF protein molecules to read or visit all the possible binding stretches of the template DNA as follows (**7**).

$$\tau_S(L) = N(\tau_{NS}+\tau_L)/L = \left(1+\varepsilon(L/L_A)^2\right)\Big/ k_f L[D];\ \tau_{NS} = 1/k_f N[D];\ \varepsilon = [P_{NS}D]/[P_B] \tag{14}$$

Here $\tau_{NS} = 1/k_f N\ [D] = 1/k_r\varepsilon$ ($s^1$) is the time required for a nonspecific collision between protein molecule and the DNA chain as a whole where the protein molecule reads on an average $L$ number of possible binding stretches within a time period of $\tau_L = L^2/6D_o$ via 1D diffusion upon each of such 3D collisions and then dissociates. Further $N/L$ is the minimum number of such 3D diffusion-mediated collisions which are required by the TF protein molecule so that it reads all the possible binding stretches of the template DNA within the time $\tau_S(L)$ and subsequently finds its specific binding site with a probability of one (scheme **SIII**, **Fig. 1A**). Contrasting from **SII**, in scheme **SIII** the 1D-diffusion length $L$ is a random variable with finite probability density function. From **Eq. 14**, one finds that the total reading time $\tau_S(L)$ will be a minimum at some critical 1D reading length $L_C$ as follows.

$$\partial_L \tau_S\left(L\right)=0;\; L_C = L_A\big/\sqrt{\varepsilon};\; \tau_{S,\min} = 2\sqrt{\varepsilon}\big/k_f\left[D\right]L_A \qquad (15)$$

In this equation $L$ will be fixed by the type of crowded environment and the average value of $L$ may be less than $L_A$. Several single molecule experiments revealed that the most probable value of that critical 1D reading length under freed environment was $L_C$ ~50-100 bps. This in turn suggests the physiological value of $\varepsilon$ ~0.4-1 which means that around ~ (30-50) % of the TF protein molecules will be always found to be nonspecifically bound with the template DNA under crowded physiological conditions. One should note that **Eq. 15** also suggest the maximum value of the enhancement factor under crowded environment as $r_{ES} = L_A\big/2\sqrt{\varepsilon}$ that is half of the value of the enhancement factor $r_{AEI}$ that is estimated in **Eq. 5** for freed environments. From **Eq. 14** one can derive the expression for the overall average bimolecular site-specific association rate ($\bar{k}_{AD}$) in the presence of dissociations of $P_{NS}D$ as follows.

$$k_{AD}\left(L\right)=k_f\beta\left(L\right);\; \beta\left(L\right)=L\Big/\left(1+\varepsilon\left(L/L_A\right)^2\right);\; \bar{k}_{AD}=k_f\int_0^N \beta\left(L\right)p\left(L\right)dL \qquad (16)$$

Here the distribution of 1D sliding lengths $L$ or its weighting function can be calculated as follows. When the residence times ($\tau$) associated (**32-33**) with the uni-molecular dissociation of single $P_{NS}D$ is distributed as an exponential then one finds $p\left(\tau\right)\propto e^{-k_r\tau}$ and subsequently we obtain the expression $p\left(L\right)\propto Le^{-\left(L/L_A\right)^2}$ that mainly originates from the fact that within the residence time $\tau$, the distance travelled by $P_{NS}D$ through 1D diffusion is $L\in\left(0,N\right)$ so that we obtain the transformation rule $\tau = L^2\big/6D_o$ . With this definition one can compute the integral in **Eq. 16** by expanding $\beta\left(L\right)$ in Maclaurin series as follows.

$$\bar{k}_{AD}=k_f\sum_{n=0}^{\infty}\left(-\varepsilon/L_A^2\right)^n\int_0^N L^{2n+1}p\left(L\right)dL;\; p\left(L\right)=2Le^{-\left(L/L_A\right)^2}\Big/L_A^2\left(1-e^{-\left(N/L_A\right)^2}\right) \qquad (17)$$

The mean ($L_\mu$) and variance ($L_\sigma$) of the possible 1D sliding lengths associated with the corresponding probability density function $p\left(L\right)$ can be derived as follows.

$$L_\mu=\left(\sqrt{\pi}L_A\mathrm{Erf}\left(N/L_A\right)-2Ne^{-\left(N/L_A\right)^2}\right)\Big/2\Omega;\; L_\sigma=\left(L_A^2\Omega-N^2e^{-\left(N/L_A\right)^2}\right)\Big/\Omega-L_\mu^2;\; \Omega=\left(1-e^{-\left(N/L_A\right)^2}\right)$$

In this equation $\mathrm{Erf}\left(z\right)=\left(2\big/\sqrt{\pi}\right)\int_0^z e^{-s^2}ds$ is the error function. When the length of template DNA is sufficiently large enough, then the mean, variance and coefficient of variation (square-root of variance/mean$^2$) of the possible 1D sliding lengths asymptotically converge to the following limits (**Fig. 2C** and **2D**).

$$\lim_{N\to\infty}L_\mu=\sqrt{\pi}L_A\big/2;\; \lim_{N\to\infty}L_\sigma=L_A^2\left(4-\pi\right)\big/4;\; \lim_{N\to\infty}\left(\sqrt{L_\sigma}\big/L_\mu\right)=\sqrt{\left(4-\pi\right)/\pi}$$

Upon evaluating the integrals under the summation in **Eq. 17** one can obtain the following expression for the overall average of the bimolecular association rate in the presence of dissociations of $P_{NS}D$.

$$\bar{k}_{AD} = k_f\left(\sqrt{NL_A}/\Omega\right)e^{-N^2/2L_A^2}S_n;\ \lim_{N\to\infty}\bar{k}_{AD} = k_f L_A\left(\pi e^{1/\varepsilon}\left(\mathrm{Erf}\left(1/\sqrt{\varepsilon}\right)-1\right)+\sqrt{\pi\varepsilon}\right)/\varepsilon^{3/2} \quad (18)$$

The sum $S_n$ in **Eq. 18** can be written in explicit form as follows.

$$S_n = 2\sum_{n=0}^{\infty}\left(\left(-\varepsilon N/L_A\right)^n \mathrm{WhittakerM}\left((2n+1)/4,(2n+3)/4,\left(N/L_A\right)^2\right)/(2n+3)\right)$$

Here the function $\mathrm{WhittakerM}(\mu,\upsilon,z)$ can be written in series form as follows (**34, 35**).

$$\mathrm{WhittakerM}(\mu,\upsilon,z) = z^{\upsilon+1/2}e^{-z/2}\sum_{s=0}^{\infty}\left(\left(\upsilon-\mu+1/2\right)_s z^s/s!\left(2\upsilon+1\right)_s\right);\ (X)_s = \Gamma(X+s)/\Gamma(X)$$

From **Eq. 18** we find that the ratio $\varepsilon$ is a critical one and the following limiting conditions exist for the overall average of the association rate and enhancement factor $r_{AD} = \left(\bar{k}_{AD}/k_f\right)$ in the presence of dissociations of $P_{NS}D$ complex.

$$\lim_{\varepsilon\to 0}\left(\lim_{N\to\infty}\bar{k}_{AD}\right) = k_f\sqrt{\pi}\,L_A/2;\lim_{\varepsilon,N\to\infty}\bar{k}_{AD} = 0;\lim_{\varepsilon\to 0}\left(\lim_{N\to\infty} r_{AD}\right) = \sqrt{\pi}\,L_A/2 \quad (19)$$

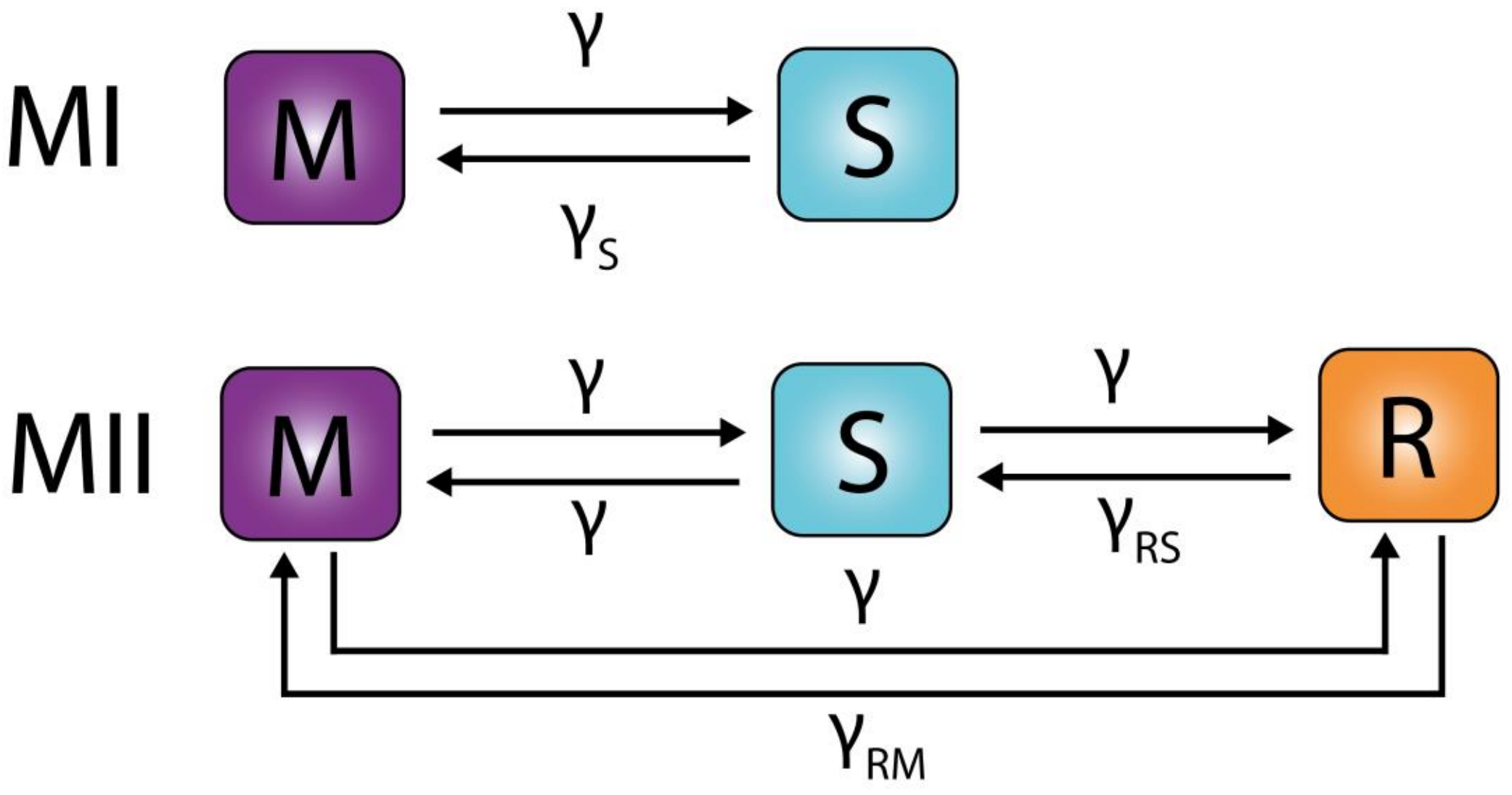

**FIGURE 3**. Various possible reading mechanisms of site-specific DNA-protein interactions. Here M is the fast-moving or +ve state, S is the slow-moving or –ve state and R is the reading state. Fast moving state is lesser sensitive to the sequence information than the slow-moving state and R is a temporal halt state where various types bonding interactions happen at the DNA-protein interface. In MI, the DBDs of DBPs fluctuates between fast and slow-moving states without a temporal pause for reading. In MII, the DBDs fluctuate between fast, slow-moving and reading states in a cyclic way whereas in MIII the fluctuations between these three states happen in a complete reversible way. The dynamical behavior of MIII will be similar to MI.

The asymptotic value of the overall bimolecular association rate given by **Eqs. 18** and **19** are indeed consistent with **Eqs. 4-9** and suggest that in the presence of dissociations of $P_{NS}D$ the rate enhancement factor is directly proportional to $L_\mu$ and $\lim_{\varepsilon\to 0} r_{AD} \approx L_\mu$ that is evident from **Figs. 2B** and **2D**. Recent experimental studies suggested (**20**) that the DBDs of DBPs fluctuate between at least two different conformational states namely fast- and slow-moving states. In the fast-moving state of DBDs, the TF protein molecules can diffuse along the DNA faster than the slow moving state. However the slow-moving state is more sensitive to the sequence information of DNA than the fast-moving state of DBDs. Efficiently tuned DNA binding domains of DBPs are the ones which are able to read the sequence information of DNA in a shortest possible timescales. Here the term ‘reading of sequence’ by DBDs denotes various intermolecular interactions such as (hydrogen) bond making/breaking which occur at the DNA-protein interface of $P_{NS}D$ complex. This means that the sequence of DNA can be read efficiently by the DBDs of DBPs when there is a transient temporal slowdown or pause in the 1D-diffusion dynamics of TF proteins at each possible binding stretch of the template DNA. In this context one can consider three possible reading mechanisms namely MI, MII and MIII as depicted in **Fig. 3**. In MI, the DBDs of DBPs fluctuates between fast and slow-moving states without a temporal pause for reading. In MII, the DBDs will fluctuate between fast, slow-moving and reading states in a parallel way whereas in MIII the fluctuations between these three states happen in a reversible way. The mechanism given in MI has already been well studied (**7, 23**) and the integral solution to the mechanism given in MIII will be similar to MI except the fact that the slow-moving state pauses at each possible binding stretch of DNA for a time period of $1/\gamma_{RM}$. The 1D-diffusion dynamics of various states of DBDs as in the mechanism MII over a DNA lattice of size $L$ bps can be described by the following set of coupled differential Chapman-Kolmogorov equations (**7, 24-26**).

$$\frac{\partial}{\partial t}\begin{bmatrix} P_M(x,t) \\ P_S(x,t) \\ P_R(x,t) \end{bmatrix} = \begin{bmatrix} -2\gamma + (D_M/2)\partial_x^2 & \gamma & \gamma_{RM} \\ \gamma & -2\gamma + (D_S/2)\partial_x^2 & \gamma_{RS} \\ \gamma & \gamma & -(\gamma_{RS}+\gamma_{RM}) \end{bmatrix} \begin{bmatrix} P_M(x,t) \\ P_S(x,t) \\ P_R(x,t) \end{bmatrix} \quad (20)$$

In this equation $D_Z$ is the 1D-diffusion coefficient associated with the conformational state $Z$ = ($M$ for fast, $S$ for slow) of DBDs, $\gamma$ ($s^{-1}$) is the flipping rate, $\gamma_{RM}$ ($s^{-1}$) is the relaxation rate constant associated with the reading state and $P_Z(x,t) = P_Z(x,t|x_0,t_0)$ are the probabilities of observing DBDs in state $Z$ = ($M$ for fast, $S$ for slow, $R$ for reading phase) at position $x$ of DNA at time $t$ starting from $x_0$ at time $t_0$. The initial and boundary conditions associated with this set of partial differential equations can be given as follows.

$$P_Z(x,t_0 \mid x_0,t_0) = \delta(x-x_0)/3;\; P_Z(0,t \mid x_0,t_0) = P_Z(L,t \mid x_0,t_0) = 0;\; Z \in (M,S,R) \quad (21)$$

Using the solutions of **Eq. 20** for the boundary conditions given **Eq. 21**, one can write down the total probability of finding DBDs at DNA position $x$ at time $t$ as follows.

$$P_T(x,t \mid x_0,t_0) = P_M(x,t \mid x_0,t_0) + P_S(x,t \mid x_0,t_0) + P_R(x,t \mid x_0,t_0) \quad (22)$$

The mean first passage time (MFPT) associated with the escape of DBDs from the lattice interval (0, $L$) obeys the following coupled backward Fokker-Planck equations.

$$\begin{bmatrix} -2\gamma + (D_M/2)d_x^2 & \gamma & \gamma_{RM} \\ \gamma & -2\gamma + (D_S/2)d_x^2 & \gamma_{RS} \\ \gamma & \gamma & -(\gamma_{RS}+\gamma_{RM}) \end{bmatrix} \begin{bmatrix} T_M(x) \\ T_S(x) \\ T_R(x) \end{bmatrix} = -\begin{bmatrix} 1/3 \\ 1/3 \\ 1/3 \end{bmatrix} \quad (23)$$

Here **Eqs. 23** can be rewritten in a simplified form as follows.

$$\begin{bmatrix} -(2-\lambda)\gamma+(D_M/2)d_x^2 & \gamma(1+\lambda) \\ \gamma(1+\rho) & -(2-\rho)\gamma+(D_S/2)d_x^2 \end{bmatrix}\begin{bmatrix} T_M(x) \\ T_S(x) \end{bmatrix} = -\begin{bmatrix} (1+\lambda)/3 \\ (1+\rho)/3 \end{bmatrix} \quad (24)$$

Here $T_R(x) = \left(\gamma\left(T_M(x)+T_S(x)\right)+1/3\right)/\left(\gamma_{RM}+\gamma_{RS}\right)$. Further we have defined the parameters $\lambda = 1/(1+\xi)$ and $\rho = \xi/(1+\xi)$ where $\xi = \gamma_{RS}/\gamma_{RM}$. Depending on the relative free energy barrier that separates the conformational transitions $R \to S$ and $R \to M$ of DBDs of DBPs one can consider the following cases.

**Case I**: $\xi = 1$; $\lambda = 1/2$; $\rho = 1/2$ where the reading state of DBDs of DBPs relaxes back either to M or S conformational states with equal probabilities or rates. Under such conditions, dynamics of the system will be independent on the transition rates ($\gamma_{RS}, \gamma_{RM}$) as follows.

$$\begin{bmatrix} -3\gamma/2+(D_M/2)d_x^2 & 3\gamma/2 \\ 3\gamma/2 & -3\gamma/2+(D_S/2)d_x^2 \end{bmatrix}\begin{bmatrix} T_M(x) \\ T_S(x) \end{bmatrix} = -\begin{bmatrix} 1/2 \\ 1/2 \end{bmatrix} \quad (25)$$

This system of **Eqs. 25** is similar to the system described by the reading mechanism **MI** where the transition rates are such that $\gamma = \gamma_S$ and $\gamma$ is multiplied by a factor of $3/2$.

**Case II**: $\xi << 1$; $\lambda \approx 1$; $\rho \approx 0$ where the reading state of DBDs of DBPs relax back to the M state more preferentially than the S state since the free-energy barrier that separates (R,S) states is much higher than the free-energy barrier that separates (R,M) states. Under such conditions, dynamics of the system will be independent on the transition rates ($\gamma_{RS}, \gamma_{RM}$) as in **Case I** as follows.

$$\begin{bmatrix} -\gamma+(D_M/2)d_x^2 & 2\gamma \\ \gamma & -2\gamma+(D_S/2)d_x^2 \end{bmatrix}\begin{bmatrix} T_M(x) \\ T_S(x) \end{bmatrix} = -\begin{bmatrix} 2/3 \\ 1/3 \end{bmatrix} \quad (26)$$

The boundary conditions for **Eqs. 23** directly follow from **Eqs. 21** as $T_Z(0) = T_Z(L) = 0$ for $Z = (M, S)$ and the overall MFPT (denoted by $\tau_{II}(x)$ where $x$ is the initial position of the DBPs on the DNA lattice) associated with the escape of DBPs from the lattice of size $L$ can be given as $\tau_{II}(x) = \sum_{Z=M,S,R} T_Z(x)$. Upon integrating **Eqs. 25** corresponding to **Case I** along with the appropriate boundary conditions, one can write the expression for the overall initial position averaged MFPT or the average time required by the DBDs of DBPs to read $L$ bps of DNA upon each 3D collision as follows.

$$\bar{\tau}_{L,II} = \int_0^L \tau_{II}(x)\,dx\Big/L = 1/6\gamma_Q + \gamma L^2/12\gamma_Q D_{AII} + \left(D_M - D_S\right)^2 \Phi_{II}(L)\Big/4\gamma_Q D_{AII}^2 \quad (27)$$

Here $\gamma_{RS} = \gamma_{RM} = \gamma_Q$ and we have defined $\bar{\Phi}_{II}(L)$ as follows.

$$\bar{\Phi}_{II}(L) = \left(1+2\left(1-\cosh\left(\sqrt{2\gamma/D_{GII}}L\right)\right)\Big/\sqrt{2\gamma/D_{GII}}L\sinh\left(\sqrt{2\gamma/D_{GII}}L\right)\right)$$

In **Eq. 25** we have defined various parameters as follows.
$D_{AII} = (D_M + D_S)/2;\ D_{GII} = D_M D_S/(D_M + D_S)$

Upon substituting the results from **Eq. 27** into **Eq. 14**, we can obtain the overall average time required by the DBDs to read the entire DNA sequence of size $N$ in the presence of conformational fluctuations in the DBDs for the mechanism SIII-MII. For large values of the flipping rate $\gamma$ one can obtain the following asymptotic approximation ($\tilde{\tau}_{S,II}$).

$$\tilde{\tau}_{S,II} \approx N\left(\tau_{NS} + 1/6\gamma_Q + \gamma L^2/12\gamma_Q D_{AII} + (D_M - D_S)^2/4\gamma_Q D_{AII}^2\right)/L \quad (28)$$

Using this asymptotic approximation, upon solving the set $\{\partial_L \tilde{\tau}_{S,II} = 0;\ \partial_{\gamma_Q}\tilde{\tau}_{S,II} = 0\}$ for the variables $(\gamma_Q, L)$ one can show that there exists an optimum conformational relaxation rate as well as 1D scanning length $(\gamma_{QC}, L_{CII})$ as follows (**Fig. 4A1-2**).

$$L_{CII} \approx \sqrt{(10D_{GII} - 7(D_M - D_S))/\gamma};\ \gamma_{QC} \approx (10D_{GII} - 7(D_M - D_S))/3\tau_{NS}(D_M + D_S) \quad (29)$$

From the determinants of the Hessian matrices $\mathrm{H}[\tilde{\tau}_{S,II}, \gamma_Q, L]$ and $\mathrm{H}[\tilde{\tau}_{S,II}, L]$ associated with the time $\tilde{\tau}_{S,\mathrm{II}}$ in **Eq. 28** evaluated at $(\gamma_{QC}, L_{CII})$, one can show that this optimum point is a saddle point in $(\gamma_Q, L)$ space as follows. Here we have defined $\Omega = (10D_M D_S - 7(D_M^2 - D_S^2))$.

$$\partial_\gamma^2 \tilde{\tau}_{S,II}\partial_L^2\tilde{\tau}_{S,II} - (\partial_{\gamma L}^2\tilde{\tau}_{S,II})^2 = -(9\gamma^2(D_M + D_S)^6 \tau_{NS}^4 N^2/\Omega^4) < 0;\ \partial_L^2\tilde{\tau}_{S,II} = (N\tau_{NS}/3L_{CII}^3) > 0.$$

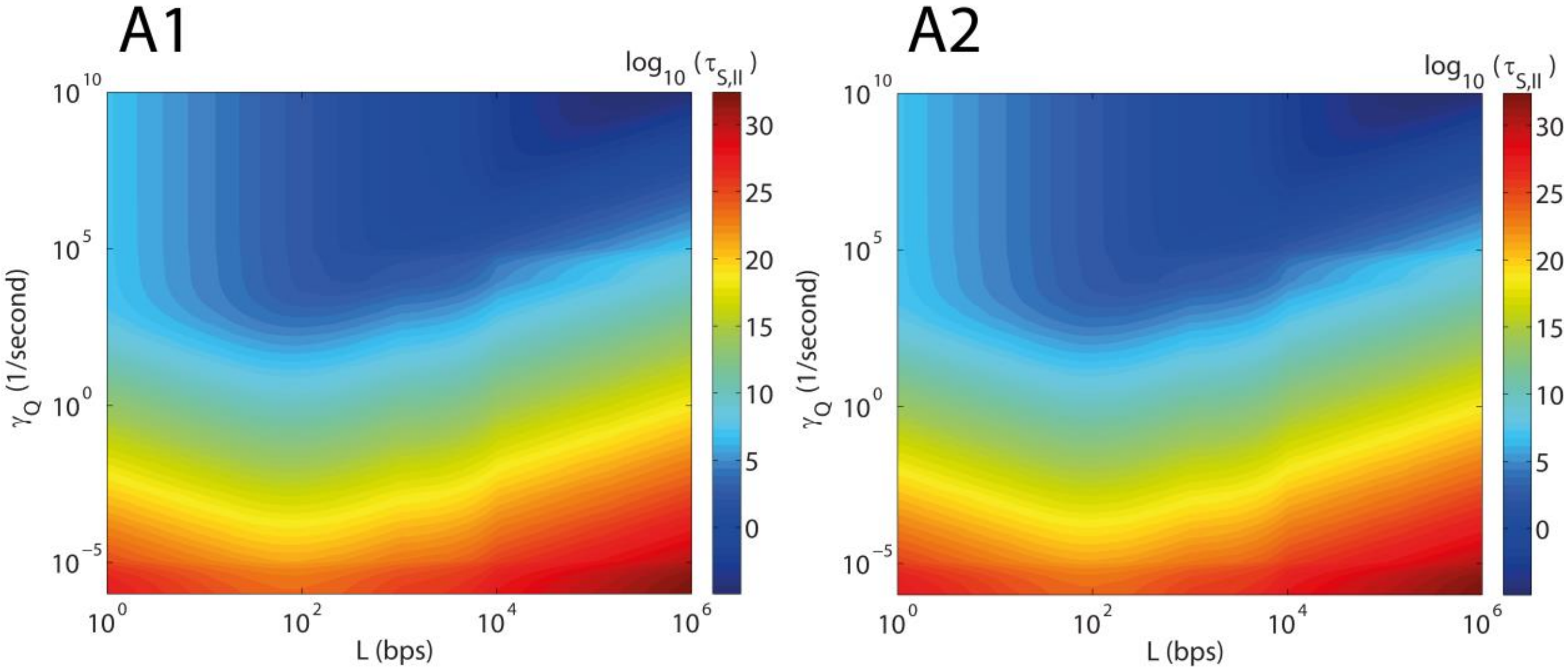

**FIGURE 4**. Variation of total time in mechanism **SIII-MII** $\tau_{S,II}$ (**A1**) that is required to read all the possible binding stretches of DNA and the asymptotic approximation $\tilde{\tau}_{S,II}$ (**A2**) with respect to changes in 1D reading length $L$ and relaxation rate $\gamma_Q$ of DNA binding domains from reading state to fast and slow moving states. Here settings for *E. coli* lac repressor system are ($D_M$ ~ 4x10$^4$, $D_S$ ~ 0.55$D_M$ as estimated from single molecule studies in **Ref. 14**) bps$^2$s$^{-1}$, $k_f$ ~ 10$^6$M$^{-1}$s$^{-1}$bps$^{-1}$(under *in vivo* conditions), $N$~4.6x10$^6$bps and $\gamma$ ~10$^1$s$^{-1}$. When we measure the concentration as number of molecules then [D] = 1 (concentration of 1 molecule inside E. coli cell will be ~1nM) and we find that $\tau_{NS}$ ~

$10^9/k_f N$. In line with **Eq. 28** we find that there exists an inflection or saddle point at $L_{CII}$ ~ 40 bps and $\gamma_{QC}$ ~ 394 $s^{-1}$.

One should note that when we denote $D_S = \omega D_M$ where $\omega \in (0,1)$ is a dimensionless quantity then the optimum points given by **Eqs. 29** will be valid only under the condition that $\left(\omega^2 + (10/7)\omega - 1\right) \geq 0$ since by definition we have $\left(\gamma_{QC}, L_{CII}\right) \geq 0$ which means that the mechanism given by **MII** will work efficiently only in the range of $\omega \in (0.5146, 1)$. When $\omega < 0.5146$ then we find that the optimum reading length increases monotonically with the relaxation rate $\gamma_Q$ as follows.

$$L_{CII} \approx \sqrt{\left(6\gamma_Q \tau_{NS}(D_M + D_S) + \left(7(D_M - D_S) - 10D_{GII}\right)\right)/\gamma}$$

The integral solution of mechanism MI can be written as follows (**7**).

$$\bar{\tau}_{L,I} = L^2/6D_{AI} + (D_M - D_S)(D_M\gamma_S - D_S\gamma)\Phi_I(L)/2(\gamma + \gamma_S)^2 D_{AI}^2 \tag{30}$$

We have defined the function $\bar{\Phi}_I(L)$ as follows.

$$\bar{\Phi}_I(L) = \left(1 + 2\left(1 - \cosh\left(\sqrt{2\gamma_S/D_{GI}}\,L\right)\right)\right/\sqrt{2\gamma_S/D_{GI}}\,L \sinh\left(\sqrt{2\gamma_S/D_{GI}}\,L\right)\Big)$$

In this equation we have defined various parameters as follows.

$$D_{AI} = (D_M + D_S\kappa_S)/(1 + \kappa_S);\ D_{GI} = D_M D_S/(D_M + D_S\kappa_S);\ \kappa_S = \gamma/\gamma_S$$

For sufficiently large value of $\gamma$ one can derive the following asymptotic approximation for the overall search time required through the mechanism SIII-MI.

$$\tilde{\tau}_{S,I} \approx N\left(\tau_{NS} + L^2/6D_{AI} + (D_M - D_S)(D_M\gamma_S - D_S\gamma)/2(\gamma + \gamma_S)^2 D_{AI}^2\right)/L \tag{31}$$

Upon solving the set $\{\partial_\gamma \tilde{\tau}_{S,I} = 0;\ \partial_{\gamma_S}\tilde{\tau}_{S,I} = 0;\ \partial_L \tilde{\tau}_{S,I} = 0\}$ for the variables $(\gamma, \gamma_S, L)$ one can obtain the critical values $(\gamma_{CI}, \gamma_{SCI}, L_{CI})$ as follows (**Fig. 5**).

$$L_{CI} \approx 2\sqrt{3\tau_{NS} D_M D_S/(D_M + D_S)};\ \gamma_{CI} \approx 3D_M/L_{CI}^2;\ \gamma_{SCI} \approx 3D_S/L_{CI}^2 \tag{32}$$

From the determinants of the Hessian matrices associated with $\tilde{\tau}_{S,I}$ evaluated at $(\gamma_{CI}, \gamma_{SCI}, L_{CI})$ one can show that thus obtained critical point is a saddle point rather than a global minimum though there is a minimum in the $(\gamma, L)$ space that is evident from the following results.

$$H\left[\tilde{\tau}_{S,I}, \gamma\right] = \left(4\tau_{NS}^3 \delta_{MS} D_S^2 N/L_{CI}\rho_{MS}^3\right) > 0;\ H\left[\tilde{\tau}_{S,I}, \gamma, L\right] = \left(2\tau_{NS}^2 N^2 \delta_{MS} D_S/9D_M^2 \rho_{MS}^2\right) > 0$$

The determinant of the three-variable Hessian matrix can be shown to be negative as follows.

$$H\left[\tilde{\tau}_{S,I}, \gamma, L, \gamma_S\right] = -\left(4N^3\delta_{MS}^2\tau_{NS}^5/9L_{CI}\rho_{MS}^4\right) < 0;\ \delta_{MS} = (D_M - D_S);\ \rho_{MS} = (D_M + D_S)$$

Similar to **Eq. 19**, using the distribution function of sliding lengths where $D_o = D_{AZ}$ one can derive expressions for the overall average association rate in the presence of conformational fluctuations in the DBDs of DBPs for cases **SIII-MI** and **SIII-MII** as follows.

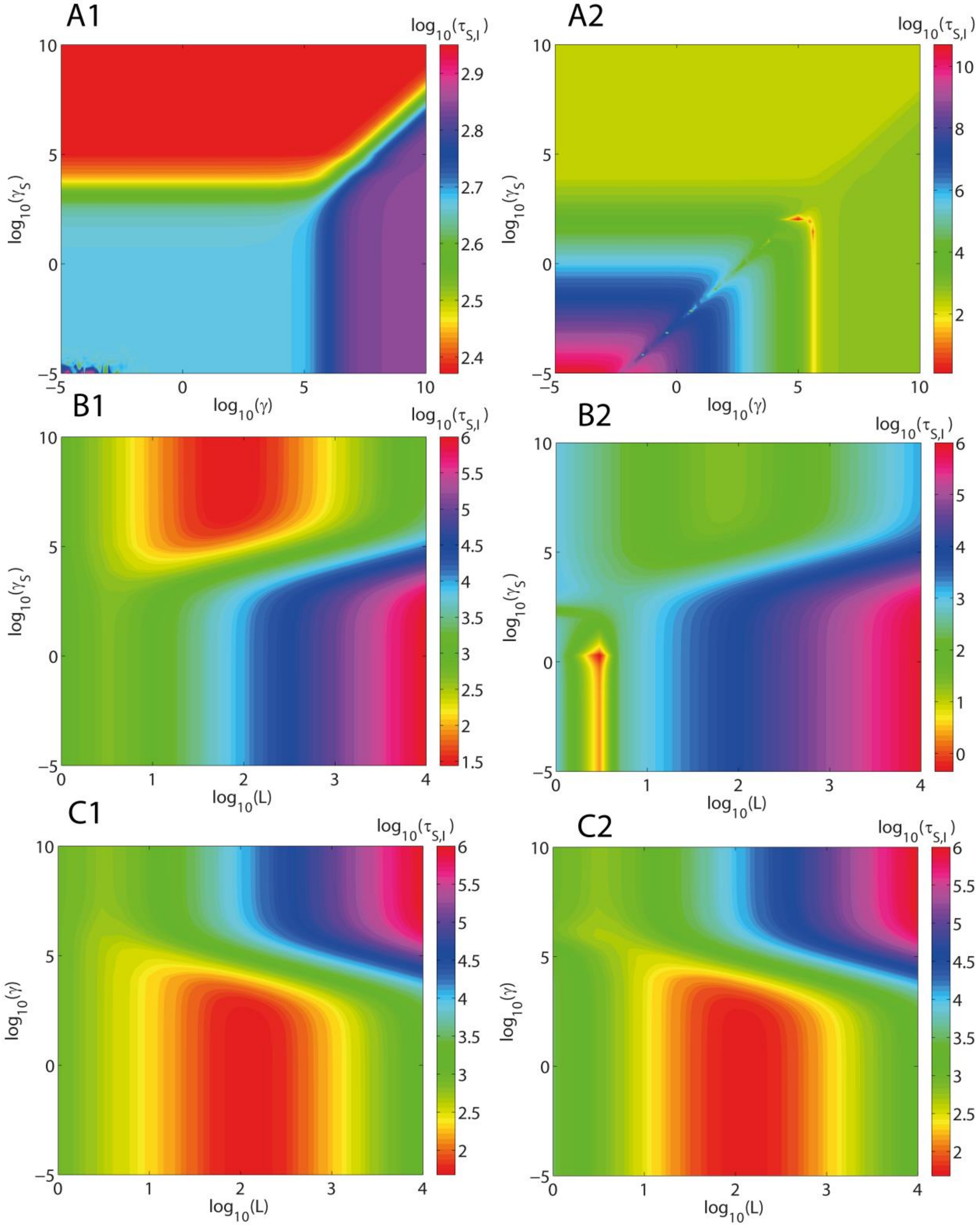

**FIGURE 5**. Variation of total search time in mechanism **SIII-MI** $\tau_{S,I}$ that is required to read all the possible binding stretches of DNA and the asymptotic approximation $\tilde{\tau}_{S,I}$ with respect to changes in 1D reading length $L$ and flipping rates ($\gamma$, $\gamma_S$) of DNA binding domains between fast and slow moving states. Here default settings for *E. coli* lac repressor system are ($D_M \sim 4\text{x}10^4$, $D_S \sim 7\text{x}10^3$ as estimated from single molecule studies in **Ref. 14**) bps$^2$s$^{-1}$, $k_f \sim 10^6$M$^{-1}$s$^{-1}$bps$^{-1}$(under *in vivo* conditions), $N$

~4.6x10$^6$ bps and ($\gamma$ ,$\gamma_S$) ~10$^4$s$^{-1}$. When we measure the concentration as number of molecules then [D] = 1 molecule and we find that $\tau_{NS}$ ~ $10^9/k_f N$. In line with **Eq. 29** we find that there exists a saddle point at $L_{CI}$ ~ 5 bps, $\gamma_{SCI}$ ~10$^3$s$^{-1}$ and $\gamma_{CI}$ ~5x10$^5$s$^{-1}$. **A1-2**: Variation of $\tau_{S,I}$ (**A1**) and $\tilde{\tau}_{S,I}$ (**A2**) with respect to changes flipping rates ($\gamma$, $\gamma_S$). **B1-2**: Variation of $\tau_{S,I}$ (**B1**) and $\tilde{\tau}_{S,I}$ (**B2**) with respect to changes flipping rates ($L$, $\gamma_S$). **C1-2**: Variation of $\tau_{S,I}$ (**C1**) and $\tilde{\tau}_{S,I}$ (**C2**) with respect to changes flipping rates ($L$, $\gamma$).

$$k_{ADZ}\left(L\right) \approx k_f L \Big/ \left(1+\varepsilon\left(L/L_{AZ}\right)^2 + \varepsilon k_r \bar{\Phi}_Z\right);\; L_{AI} = \sqrt{6D_{AI}/k_r};\; L_{AII} = \sqrt{12\gamma_Q D_{AII}/\gamma k_r} \quad (33)$$

The functions $\bar{\Phi}_Z$ where the subscript $Z$ = ($I$, $II$) are defined as follows.

$$\tilde{\Phi}_I = \left(D_M - D_S\right)\left(D_M\gamma_S - D_S\gamma\right)\Big/2\left(\gamma+\gamma_S\right)^2 D_{AI}^2;\; \tilde{\Phi}_{II} = 1\big/6\gamma_Q + \left(D_M - D_S\right)^2\Big/4\gamma_Q D_{AII}^2$$

Similar to the **Eq. 18** using the respective probability density function $p_Z\left(L\right)$ one can derive an expression for the average of the overall site-specific association rate in the presence of dissociations from **Eq. 31** as follows.

$$\bar{k}_{ADZ} = \int_0^N k_{ADZ}\left(L\right) p_Z\left(L\right) dL;\; r_{ADZ} = \bar{k}_{ADZ}\big/k_f;\; p_Z\left(L\right) = 2Le^{-\left(L/L_Z\right)^2}\Big/L_Z^2\left(1-e^{-\left(N/L_Z\right)^2}\right) \quad (34)$$

Using this equation one can derive the following limiting values of the overall average site-specific association rate.

$$\lim_{\varepsilon\to 0}\left(\lim_{N\to\infty} \bar{k}_{ADZ} = k_f L_{AZ}\left(\pi e^{w/\varepsilon}\sqrt{w/\varepsilon}\left(\mathrm{Erf}\left(\sqrt{w/\varepsilon}\right)-1\right)+\sqrt{\pi\varepsilon}\right)\Big/\varepsilon^{3/2}\right) = k_f\sqrt{\pi}L_{AZ}\big/2 \quad (35)$$

Here $w = \left(\tilde{\Phi}_Z k_r \varepsilon + 1\right)$. **Eq. 19** and **Eq. 35** are the central results of this paper which connect the maximum possible value of the rate enhancing factor that can be achieved in site-specific DNA-protein interactions with respect to increase in size of the template DNA length $N$ and the maximum possible average 1D sliding lengths ($L_A$ and $L_\mu$).

## RESULTS AND DISCUSSION

Site-specific binding of transcription factor proteins with *cis*-regulatory motifs (CRMs) of the promoters is critical for activating various genes particularly in eukaryotes. Understanding the precise mechanism of site-specific DNA-protein interactions is crucial to understand and unknot the dynamical aspects of gene regulation. TFs can bind with the corresponding CRMs via either pure 3D diffusion or combination of 1D- and 3D-diffusion. Our current results and earlier studies clearly ruled out the possibility of 3D-only model. At this point it is necessary to explain how exactly the rate enhancement happens in 1D3D model. In 3D-only model, TF protein reads only one possible binding stretch upon each 3D-diffusion mediated collision with the template DNA that requires ($\tau_{NS}$ ~ $1/k_f$ [D] $N$) amount of time where we have defined the unit of $k_f$ as M$^{-1}$s$^{-1}$bps$^{-1}$. Whereas in 1D3D model, the TF protein reads $L$ number of possible binding stretches upon each 3D diffusion-mediated collision that requires ($\tau_L$ + $\tau_{NS}$) amount of time where $\tau_L$ = $L^2/6D_o$. With this framework the TF protein can visit all the possible binding stretches of template DNA within a time of $\tau_S$ ($L$) = $N$ ($\tau_L$ + $\tau_{NS}$)/$L$ as given in **Eq. 14**. When there are no other roadblock protein molecules on the 1D-diffusion path of TF protein of interest then the TF protein of interest can read on an average $L_A$ number of possible binding stretches upon each 3D-diffusion mediated collisions with DNA. That is the reason why the overall enhancement factor is directly proportional to the average 1D sliding-length $L_A$ or 1D mean free path length $L_M$ when $L_M < L_A$. **Fig. 2A** demonstrates the asymptotic behavior of the rate enhancement factor $r_{AEI}$ with respect to changes in the dissociation rate constant $k_r$ and size of the template DNA $N$ as given in **Eq. 5** where we had set $D_o \sim 10^6\,\mathrm{bps^2 s^{-1}}$ (**14**, **15**). From **Fig. 2B** one can conclude that scheme **SII** overestimates the maximum possible value of the overall rate enhancement factor ($r_{AEI}$) than **SIII** ($r_{AD}$). This is mainly because in the calculation of $k_L$ in **SII** we have assumed that the nonspecific DNA-protein complex $P_{NS}D$ does not dissociate until it reaches the specific binding site which is not true in most of the crowded *in vitro* as well as *in vivo* physiological conditions. We have considered three possible reading mechanisms of DBDs of DBPs as depicted in **Fig. 3**. Upon analyzing the variation of the overall search times ($\tau_{S,I}, \tau_{S,II}$) required by the non-specifically bound TF protein to scan the entire DNA chain via combination of 1D and 3D routes with respect to changes in sliding lengths ($L$) and conformational flipping rates $(\gamma, \gamma_S)$ of DNA binding domains we found that out of reading mechanisms MI, MII and MIII the critical values obtained for the 1D diffusion mediated reading length $L_{CZ}$ corresponding to $Z = II$ (MII) as shown in **Figs. 4** and **5** is closer to the probable value of 1D sliding length $L_A$~60bps that corresponds to the observed overall rate enhancement factor $r_{AEI}$ ~ 100 bps (**Eq. 8**). This result suggests that **SIII-MII** will be the more plausible as well as efficient mechanism of reading the sequence information of DNA than mechanisms given in MI and MIII. The maximum achievable value of the overall rate enhancement factor in **SIII-MII** can be written explicitly from **Eq. 35** as follows.

$$\lim_{\varepsilon \to 0}\left(\lim_{N\to\infty} r_{ADII}\right) = \sqrt{3\pi\gamma_Q\left(D_M + D_S\right)/\gamma k_r} \tag{36}$$

In this equation the dimensionless ratio ($\gamma_Q/\gamma$) acts as tuning parameter associated with the overall rate enhancement factor in **SIII-MII** mechanism. Similarly for **SII-MII** mechanism the parameter $\kappa_S$ can act as a tuning parameter. Several models (**23**) considered the template DNA as an intact sphere or cylinder like molecular object and overestimated the non-specific collision time $\tau_{NS}$ as ($\tau_{NS}$ ~ $1/k_{fa}$ [D] where the unit of $k_{fa}$ is $M^{-1}s^{-1}$) that introduced an unrealistic argument that the 1D-diffusion always retards the overall searching dynamics in 3D1D model. The ratio $N/L$ in the expression of the search time in **SIII** $\tau_S$ ($L$) is the minimum number of 3D diffusion-mediated collisions between the DNA template and TF protein required by the TF molecule to read all the possible binding stretches of DNA so that the probability of finding the specific-site by the protein will be one. In this context Zhou in reference (**23**) considered the ratio $\eta = L/N$ in the definition of $\tau_S$ ($L$) as the probability of finding the specific-patch on DNA cylinder and he argued that it should be a position dependent quantity since the possible value of $L$ is dependent on the DNA sequence and further he muddled $\eta$ with function $\phi(L)$ of **Eq. 3** since both of these quantities are defined in totally different contexts. In fact the quantity $L$ in **Eq. 3** (**SII**) is the distance (fixed) between the specific binding-site and the initial landing position of the TF molecule on DNA via 3D collision whereas in the definition of $\tau_S$ ($L$) it is the minimum number of possible binding stretches that is supposed to be read by the TF molecule upon each 3D diffusion-mediated collision with the template DNA. This means that $\eta$ is not the same as that of $\phi(L)$ since their definitions are valid only under respective circumstances. Particularly the definition of $\phi(L)$ assumes that the nonspecifically bound TF protein will not dissociate from DNA until it finds the specific binding site whereas $\eta$ assumes that the nonspecifically bound TF protein molecules dissociate from template DNA and then associate back several times while reading $L$ number of possible binding stretches.

## CONCLUSIONS

We have developed revised theoretical concepts on how transcription factor proteins locate their specific binding sites on DNA faster than the three-dimensional (3D) diffusion controlled collision rate limit. We have shown that the 3D-diffusion controlled rate limit can be enhanced only when the protein molecule reads several possible binding stretches of the template DNA via one-dimensional (1D) diffusion upon each 3D-diffusion mediated collision or non-specific binding event. The overall enhancement of site-specific association rate seems to be directly proportional to the maximum possible sliding length ($L_A$, square root of ($6D_o/k_r$) where $D_o$ is the 1D-diffusion coefficient and $k_r$ is the dissociation rate constant connected with the nonspecific DNA-protein complex) associated with the 1D diffusion of protein molecule on DNA. Upon considering several possible mechanisms of reading the sequence of DNA by the protein molecules, we have found that the DNA binding proteins efficiently locate their cognate sites on DNA by switching across fast-moving, slow-moving and reading states of their DNA binding domains (DBDs) in a cyclic manner. It seems that the overall rate enhancement factor asymptotically approaches a limiting value which is directly proportional to the maximum possible 1D sliding length irrespective of the type of mechanism of reading as the total length of the DNA chain that contains the cognate site increases. These results are consistent with the *in vitro* experimental observations.